# Long Range Coupling between Metallic Nanocavities

Adi Salomon[1,2] and Yehiam Prior[1]

[1]Department of chemical physics, Weizmann Institute of Science, Rehovot, Israel
[2]Department of Chemistry, Institute of Nanotechnology and Advanced Materials, Bar Ilan University, Raman Gan, Israel

Michael Fedoruk[3] and Jochen Feldmann[3]

[3]Department of Physics, Ludwig-Maximilians-Universtität München, Munich, Germany

Radoslaw Kolkowski[4,5] and Joseph Zyss[5]

[4]Laboratoire de Photonique Quantique et Moléculaire, Institut d'Alembert, Ecole Normale Supérieure de Cachan, France
[5]Institute of Physical and Theoretical Chemistry, Faculty of Chemistry, Wroclaw University of Technology, Poland

When two or more metallic nanoparticles are in close proximity, their plasmonic modes may interact through the near field, leading to additional resonances of the coupled system or to shifts of their resonant frequencies. [1-10] This process is analogous to atom-hybridization, as had been proposed by Gersten and Nitzan[11] and modeled by Nordlander et al.[9]. The coupling between plasmonic modes can be in-phase (symmetric) or out-of-phase (anti-symmetric), reflecting correspondingly, the 'bonding' and 'anti-bonding' nature of such configurations[12-14]. Since the incoming light redistributes the charge distribution around the metallic nanoparticles, its polarization features play a major role in the nonlinear optical probing of the energy-level landscape upon hybridization [1, 15]. Thus, controlling the nature of coupling between metallic nanostructures is of a great importance as it enables tuning their spectral responses leading to novel devices which may surpass the diffraction limit.

Nanocavities (nanoholes carved in a thin metal film) can be viewed as complementary to nanoparticles and their respective linear optical responses are related by Babinet's principle[16]. However, when the interaction between neighboring structures is discussed, a fundamental difference exists between nanocavities and nanoparticles. The interaction between individual nanoparticles decays rapidly (typically within tens of nanometers). However, in the case of nanocavities, propagating surface plasmons polaritons (SPP), which are supported by the metallic film between the nanocavities, provide a channel for long-range coupling between the nanocavities, and therefore, fundamentally changing the interaction picture. This additional coupling mechanism offers a way to modulate the linear and nonlinear responses of such nanostructures and may lead to a strong amplification of the observed signal.

The coupling between metallic nanostructures (nanocavities or nanoparticles), in particular its polarization properties, can be ideally probed by nonlinear optical measurements. In particular, , second harmonic generation (SHG) [17, 18] offers two benefits: it is the lowest order nonlinear optical mechanism and it is highly sensitive to symmetry disruption as opposed to linear or cubic processes, which are less prone to reflect symmetry changes. More specifically, the dependence of SHG responses on the polarization of the incident beam provides a direct information about the properties of the second order susceptibility tensor $\chi_{ijk}^{(2)}$ which critically depends on the structure symmetry and the charge distribution at the metallic nanostructure surface. [19-21] Being a coherent process, SHG, can provide important information about the nature of interaction between participating oscillating dipoles. More explicitly, if all the dipoles emanating from the metallic nanostructures are in phase, we expect high directionality of the SH emission. In addition, the SH intensity scales with $N^2$, where N is the number of the metallic nanostructures and therefore when all are coherently emitting, the signal is highly amplified. An incoherent process, on the other hand, scales linearly with the number of participating entities. However, generation of a coherent SH response from an ensemble of metallic nanostructures (particles or holes) is a challenging task. One of the reasons is that the SH responses arising from surface defects or the host material may overwhelm the intrinsic signal of the metallic nanostructures, eventually resulting in phase randomization of the emitted dipoles and relatively low signals. Thus, the overall SH emission from an ensemble of metallic nano structures usually refers to the incoherent process rather than to the coherent one.

Herein we report on the coherent coupling between plasmonic modes of triangular cavities, which are arranged in a line and are separated by hundreds of nm. We compare their nonlinear response to that of a complementary row of triangular nanoparticles of identical shape, size and structure. The interaction is characterized by measuring the SHG response at different polarizations of the input beam. This NLO based approach is further complemented by additional linear dark field measurements. We show that while coupling between the nanocavities induces a strong anisotropy in the SHG emission, such is not the case for nanoparticles. For identical geometrical parameters, no coupling is observed in the case of nanoparticles, thus breaking the often postulated complementary between cavities and particles.

We chose to concentrate on equilateral triangular nanostructures, cavities or particles. [22] This choice leads to a striking difference between the SHG responses of an individual triangle which has a threefold symmetry, as compared to a chain of three triangular structures, abiding to a lower symmetry only when coupled. Without coupling, their behavior mimics that of an individual triangular structure.

The samples consist of equilateral triangular nanocavities drilled in a thin silver film and their complementary structures, triangular silver nanoparticles, all of a typical side length of 200nm (see figures 1a,b). The triangular cavities/particles are arranged in a linear configuration separated by a distance of about 400 nm (figures 2a,e).

We first study the nonlinear (SHG) response of individual triangular nanostructures (particle and cavity) as a function of the polarization of the incoming beam. The results are presented in Figure 1. The total SHG emission intensity patterns for both individual triangular nanoparticle and nanocavity structures are barely dependent on the input beam polarizations. Thus no directionality of the SHG emission is observed and the complementary structures display a similar quasi-isotropic behavior [23] .

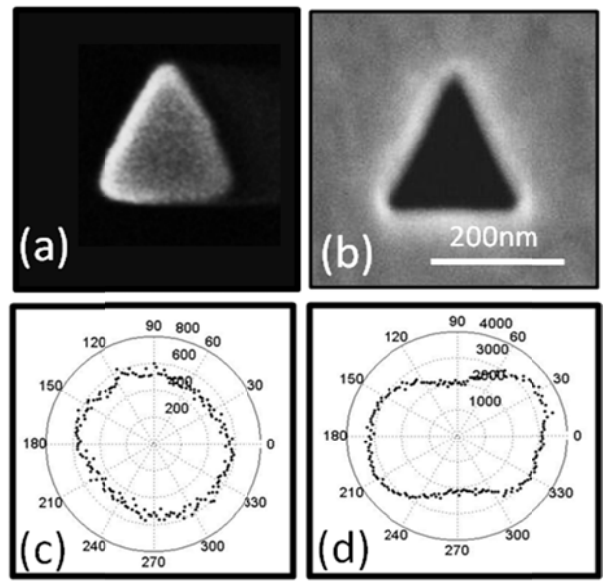

**Figure 1:   SEM images of triangular nanoparticle and nanocavity as complementary nanostructures and their corresponding SHG emission patterns.**   (a) A triangular nanoparticle with side length of ~200nm.  (b) A triangular nanocavity with the same geometrical dimensions. (c-d) are the corresponding polar plots of the SHG total emission patterns as a function of the polarization of the incoming beam.

Next we study the SHG response of a linear chain of three triangular nanostructures. If these nanostructures are coupled and behave as a single entity, the global symmetry of the assembly governing their overall response is then reduced and the SHG response and its emission patterns should change accordingly. Figure 2 depicts the spatial distribution of the SHG responses at two orthogonal polarizations for the two complementary nanostructures, triangular cavities and particles respectively. The spatial distribution of the SHG intensity under those two orthogonal polarizations for the **nanoparticles** is shown in Figures.2b & 2c.  In both cases the intensity seems to be almost uniformly spread, and there is very little difference between excitations by the two orthogonal incoming polarizations. However, for **nanocavities** (Figures 2f & 2g), a dramatic change in the SHG responses is observed depending on whether the input polarization is pointing along the common two-fold (Y-axis) or perpendicular to it (X-axis). When the input polarization is along their common axis, not only is the SHG intensity much higher, but also it is spatially concentrated at the middle triangle, thus evidencing a different behavior for the system.

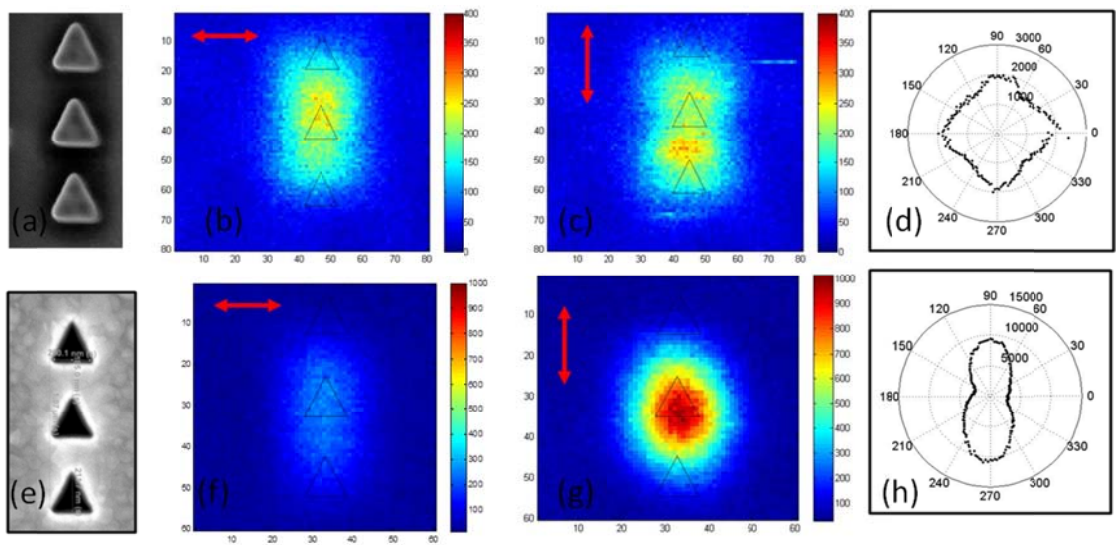

**Figure 2: SHG responses as a function of the polarization of the incoming beam for triplets of triangular nanoparticles and nanocavities**. (a) SEM image of a triplet of triangular nanoparticles with side length of 200nm and separation of ~400nm, and their complementary nanocavities structures (e). Scanned image of the SHG emission of **nanoparticles** with incident polarization along the X-axis (b), and along the Y-axis(c) and their corresponding total emission(d). Corresponding scanned image of the SHG emission of **nanocavities** with incident polarization along the X and Y axes (f and g) and their corresponding total emission (h), showing a highly anisotropic emission pattern. The fundamental beam wavelength was 940nm for both cases.

To further characterize the polarization properties of such structures, we measured their complete dependencies on the input beam polarization angles (from 0° to 360°). The nanoparticles (figure 2d) display an almost uniform signature, with similar intensities for all input polarization directions as for an isolated triangular particle or cavity. However, the nanocavities (figure 2h) display a very different behavior and a clearly anisotropic dipolar emission is observed, revealing strong coupling between the plasmonic modes localized on cavities. Moreover, a higher emission level (by a factor of 4-5) is observed when the triangular cavities are excited by light polarized along the common Y-axis as compared to the X- axis.

In addition, the anisotropy is demonstrated in a nice pictorial way by scanning the SHG response of 'coupled' cavities arranged along a circle, pointing towards the center. Figure 3 shows the change in the SHG response when the polarization of the incoming beam is set along the X-axis. The SHG response clearly depends on the orientation (angle) of the 'coupled' cavities with respect to the x-axis, indicating stronger coupling between the three cavities when properly aligned.

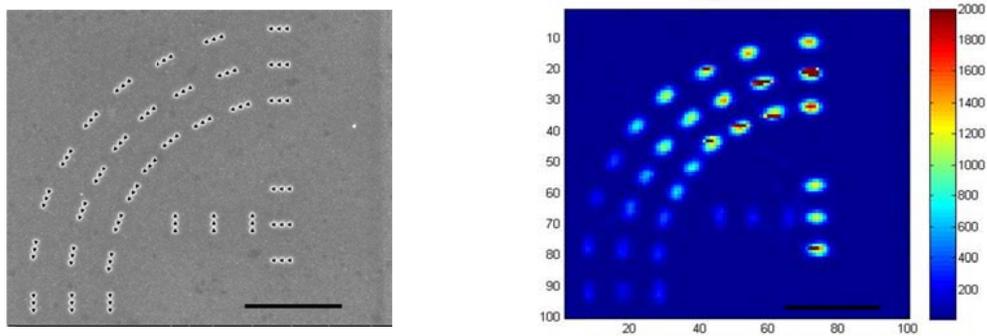

**Figure 3: SHG responses of the triangular nanocavities triplets for x-axis polarization of the incoming beam**. (a) SEM image of a triplet of triangular nanoparticles with side length of 200nm and separation of ~400nm arranged in a rainbow like configuration (b). A scanned image of SHG emission with incident polarization along the X-axi with the suppressed/enhanced SHG clearly observed. The fundamental beam wavelength was 940nm for both cases, and the scale bar is 5 micron.

In order to confirm that a pure SHG signal was observed rather than any contributions of luminescence or fluorescence [24], the spectra of the emitted light was measured as shown in Figure 4. For both nanoparticles and nanocavities, a narrow linewidth signal peaking exactly at the second-harmonic frequency is the only component present in the spectra, a clear and unambiguous spectral signature of SHG. Figure 4a shows the SHG spectrum observed from triangular nanoparticles excited by two orthogonal polarizations. In this case, both polarizations exhibit similar intensities in agreement with Figure 2. In the case of the nanocavities, however, the SHG intensity (Figure. 4b) is higher by factor ~5 when excited along the common axis (Y-axis) compared to that of the X-axis, in agreement with the results already shown.

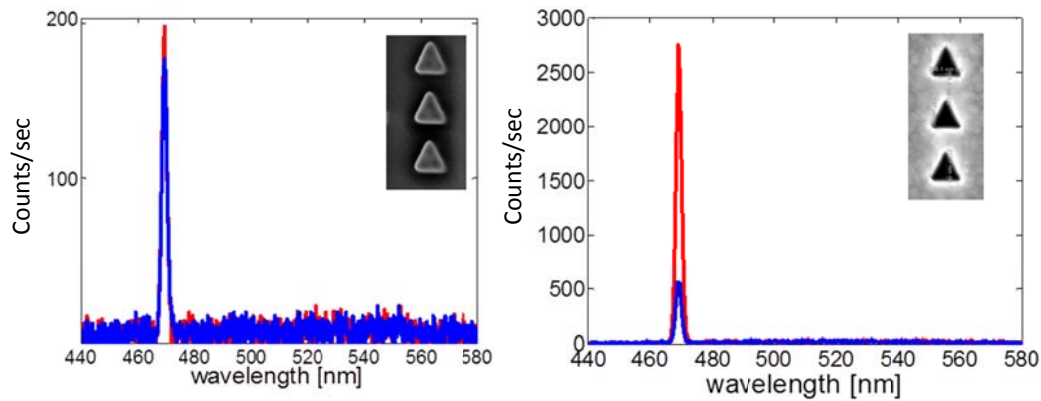

**Figure 4: SHG emission spectra when excited at 940nm fundamental wavelength** (a) Spectrum of a triplet of triangular nanoparticles; blue curve, incoming beam polarized along the X axis, red – incoming beam polarized along the Y axis. (b) Spectrum of a triplet of triangular nanocavities; blue curve, incoming beam polarized along the X axis, red – incoming beam polarized along the Y axis. See text for a discussion of the results.

The observed SHG response of the two complementary structures indicates that different physical mechanisms are involved in each case. While the three nanoparticles respond as individual particles where no interaction is coupling them, the three nanocavities respond as a single entity with a much stronger, spatially concentrated SHG emission. The original three-fold symmetry of the triangular cavity is then lowered, resulting in an anisotropic dipolar SHG emission pattern, where all the cavities are in-phase and respond at unison as a single dipole.

To further confirm coupling between the nanocavities, linear scattering measurements were performed by dark field microscopy. Figure 5 depicts the main observations. For individual triangular cavities the plasmonic mode appears around 800nm, and is insensitive to the polarization of the incoming light. However, when a triplet of cavities is measured, the spectrum is very sensitive to the polarization of the incoming beam. When the incident light is polarized along the common triplet axis, two main peaks are observed, one at about 905nm and the other at about 775nm. These two peaks may reflect the σ and σ* states in the hybridization model, where the more intense red-shifted mode can be identified as the in-phase σ mode, while the less intense blue-shifted mode may be identified as the σ* out-of-phase mode. Note that as described by the hybridization model both modes are observed.[1, 25] When the triplet of cavities is probed by light that is polarized perpendicular to the triplet axis we observed a red-shifted peak at about 875nm, which is probably the out-of-phase π mode . The π* mode is not resolved in these measurements.

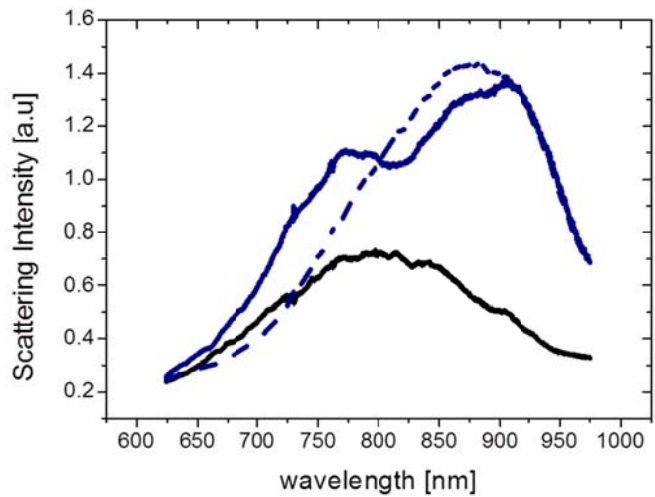

**Figure 5**: **Scattering spectra obtained by dark field microscopy for isolated triangular holes (black line), and a triplet of triangular nanocavities (solid and dashed blue lines)**. The solid blue line depicts the measurement for an input polarization along the triplet axis, the dashed blue line is for an input polarization perpendicular to the triplet axis. Note that at 940 nm, the wavelength used in our nonlinear experiments, the scattering from the triplet intensity for both polarizations is identical.

The measured linear spectra confirm previously reported measurements on nanoparticles, whereby strong coupling leads to splitting and shifts of the spectral lines. It supports our conclusion of strong coupling between the plasmonic modes of these triangular cavities. Furthermore, it reaffirm our observation of the long range coupling for nanocavities as opposed to the known short range plasmonic coupling of metallic nanoparticles. Unlike the case of nanoparticles, we are able to probe here also the so-called dark modes (σ*), which is slightly blue shifted with respect to individual modes. The single peak which is observed for the orthogonal polarization is red shifted rather than blue shifted as already observed for nanoparticles.

When it comes to coupling between neighboring metallic nanostructures, a fundamental difference exists between nanocavities and nanoparticles and they cannot be considered as complementary. Unlike nanoparticles, nanocavities may couple through excitation of SPP propagating at the metallic film. [26] These propagating SPP modes connect the localized surface plasmons (LSP) at neighboring nanocavities, thus inducing strong mutual interactions. The degree of coupling depends on the SPP excitation wavelength and its propagation length and therefore on the distance between the cavities. The intrinsic parameters of the metallic film itself play a major role as well. For example, poor coupling was observed (not shown here) when the metallic surface between the cavities was rough leading to scattering of the SPP modes, phase randomization of the dipoles, and therefore reduced coupling efficiency between the LSP modes. A high fabrication quality of the silver film in terms of roughness and grain size, as well as of the structures themselves, proved to be essential for our current observations.

*Estimation of the coupling degree*

The coupling degree between the cavities and their mutual interaction is subjected to several parameters as already mentioned above. The amount of coupling can be quantitatively inferred from the SHG emission analyzed along two orthogonal angles (X and Y) for a range of input beam linear polarizations spanning by rotation all directions within the plane of the sample.

In what follows we perform an analysis of the symmetry properties of such structures as measured by the SHG, and introduce a tensor-symmetry-based model to analyze the degree of coupling between the nanostructures in connection with their symmetry lowering when going from uncoupled to coupled nano-cavities.

To a good approximation, our samples can be viewed as two-dimensional, and contributions from the Z component (perpendicular to the plane of the thin film) of the electric fields can be neglected. The dominant contributions to the susceptibility tensor $\chi_{ijk}^{(2)}$ then span only the Y and X directions, leaving $\chi_{YYY}^{(2)}$, $\chi_{YXX}^{(2)}$, $\chi_{XXX}^{(2)}$ and $\chi_{XYY}^{(2)}$ as the only four independent coefficients. The presence of an additional mirror plane normal to the X axis further leads to the cancellation of tensor elements which are anti-symmetric with respect to this mirror symmetry. Thus, terms where the X index makes an odd number of occurrences vanish, namely $\chi_{XXX}^{(2)} = \chi_{XYY}^{(2)} = 0$. For a higher symmetry corresponding to three-fold rotational invariance around Z, such as for equilateral triangular structures, the two remaining coefficients are further linked by a tensor relation which expresses the cancellation of all dipolar-like physical entities [27], namely:

$$\chi_{YYY}^{(2)} + \chi_{YXX}^{(2)} = 0 \qquad (1)$$

which leaves only one independent coefficient .

Such is the case for individual nano-cavities, or a non-interacting set of these (likewise for nano-particles).

However, when coupling is coming into play, the overall symmetry is reduced to a lower level, relation (1) no longer holds, and the $\chi_{YYY}^{(2)}$ and $\chi_{YXX}^{(2)}$ terms become then independent.

According to these general symmetry considerations, the level of coupling between the nanocavities can be monitored via the dependence of the emitted SHG signal on the input beam polarization. As can be seen in Figure 2, for a significant level of coupling, when the input polarization is along the Y-axis, the three cavities respond collectively (Figure 2g) and the polarization signature of the emitted light (figure 2h) changes from a four-fold octupolar shape, characteristic of three-fold symmetry, [28] [22, 29] to an elongated dipolar signature characteristic of a reduced, two-fold reflection symmetry. Thus, we define a nonlinear anisotropy coefficient ρ, which reflects the departure from three-fold symmetry due to the coupling between the nanocavities, according to:

$$\rho = \chi_{YXX}^{(2)} / \chi_{YYY}^{(2)} \qquad (2)$$

The ρ parameter accounts for nonlinear anisotropy and enables to quantitatively monitor the transition from three independent objects to a single, internally coupled entity of lower symmetry. As the level of coupling between individual nanostructures increases, $\chi_{YYY}^{(2)}$ tends to dominate over $\chi_{YXX}^{(2)}$ and ρ ultimately approaches 0. On the other hand, in the absence of coupling between the individual nano-structures, the system maintains its original three fold symmetry, relation (1) holds and ρ=−1. These two extreme cases of individual contributions ($\rho = -1$) and the fully coupled one ($\rho = 0$), calculated under realistic experimental conditions, are shown and fully discussed in the supplementary material.

To investigate the degree of coupling between the triangular nanocavities, we performed detailed polarization analysis of the individual X- and Y components of the emitted SHG radiation as shown in figure 6. A dipolar-like pattern is observed for the polarization along the Y-axis (blue curve); while an octupolar pattern of weaker intensity is observed along the X-axis (red curve exhibiting a butterfly-like shape). The dipolar pattern results from the reduced symmetry due to the strong coupling between the triangular cavities. However, the coupling is not fully accomplished and the X component emission pattern features a residual octupolar pattern [22, 29]. Within the model described above, the data was best fitted (other than intensity scaling, $\rho$ is the only fitting parameter) to $\rho = -0.5$. This value indicates a strong, but less than full level of coupling between the nanocavities. We note that under identical conditions, the triangular nanoparticles are not coupled at all, thus maintaining the clearly octupolar signature typical of uncoupled triangles (figure2d).

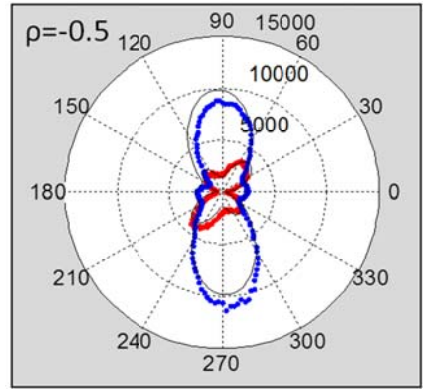

**Figure 6: Fitting the experimental polar plots of the coupled system to the model**. Blue and red dots are the experimentally observed data points (Y and X components respectively) for a triplet of triangular nanocavities. The black curve is our theoretical model calculated for ρ = -0.5, indicating a strong, but less than full, coupling. Details of the calculation are given in the supplementary material.

In conclusion, long range coupling between three triangular nanocavities is observed by SHG, in strong contrast with the case of nanoparticles with the same geometrical parameters. Complementary dark field scattering microscopy experiments confirm the existence of a significant coupling. We explain that the coupling is mediated by surface plasmons propagating in the metal film, somwthing that is not possible for the case of nanoparticles which couple through air or the dielectric layer. Increasing the coupling degree between the cavities may be achieved by optimizing the distance between them or/and the excitation wavelength of the input beam. A full understanding of long range coupling between neighboring nanostructures will prove essential for introduction specific structures with unique nonlinear optical properties.

**Experimental:**

The metallic nanostructures (cavities and particles) were fabricated by a focused ion beam machine (*FEI, Helios Nano Lab 600i*) in a 200nm thick silver film (see figs 1&2). The film was evaporated onto a clean fused silica glass under high vacuum conditions; its roughness and grain size were measured by Atomic Force Microscopy and Scanning

Electron Microscopy to be smaller than 1nm and 50nm respectively. In order to have similar indices of refraction on both sides of the nanocavities or around the nanoparticles, the silver surfaces were covered by a 150 nm layer of polyvinyl alcohol (PVA), with an average refractive index similar to that of glass (~ 1.5) in the relevant wavelength range of the visible to near infrared. The sample was mounted on a piezo stage and was scanned with a resolution of 40nm at the focal plane of a focused beam from a Ti:Sapphire laser (*Spectra-Physics Mai-Tai HP*, 100 fsec, 80MHz), 2-6 mWatt at the entrance lens, with a fundamental incoming beam tunable between 750nm-980nm. Most of the experiments reported here were performed at an input laser wavelength of 940 nm. The main reason for selecting this wavelength is the fact that the silver film supports plasmon propagation at both the fundamental and at the second harmonic frequency of 470 nm. The laser was focused through the glass substrate using a 0.7 NA objective (×60), resulting in a spot size of about 0.9 µm. The SH signal was collected in reflection mode through the same objective, and directed to two avalanche photodiodes (APD, PerkinElmer) that measure the SH intensity along the X and Y perpendicular polarization directions. A dichroic mirror was used to block the reflected fundamental beam, and appropriate band-pass filters (*Semrock*) were used to further isolate the SH radiation. The SHG polarization dependence measurements have been performed by rotating the linear input beam polarization using a half wave plate and a polarizer (see figure s1 in supplementary materilas for set-up details).

**Acknowledgements:** We acknowledge the support of the Weizmann-CNRS NaBi LEA laboratory. This work was funded, in part, by the Israel Science Foundation, grant no. 1242/12 and by the Leona M. and Harry B. Helmsley Charitable Trust. RK acknowledges the government for the Polish-French "co-tutelle" PhD program. JZ acknowledges support of the Humboldt Foundation for visits to LMU.

# Long Range Coupling between Metallic Nanocavities

Adi Salomon[1,2] and Yehiam Prior[1]

[1]Department of chemical physics, Weizmann Institute of Science, Rehovot, Israel
[2]Department of Chemistry, Institute of Nanotechnology and advanced materials, Bar Ilan University, Raman Gan, Israel

Michael Fedoruk[3] and Jochen Feldmann[3]

[3]Department of Physics, Ludwig-Maximilians-Universtität München, Munich, Germany

Radoslaw Kolkowski[4,5] and Joseph Zyss[5]

[4(]Laboratoire de Photonique Quantique et Moleculaire, Institut d'Alembert, Ecole Normale Supérieure de Cachan, France
[5]Institute of Physical and Theoretical Chemistry, Faculty of Chemistry, Wroclaw University of Technology, Poland

## Supplementary Information

The experimental optical set-up is shown in Fig. S1

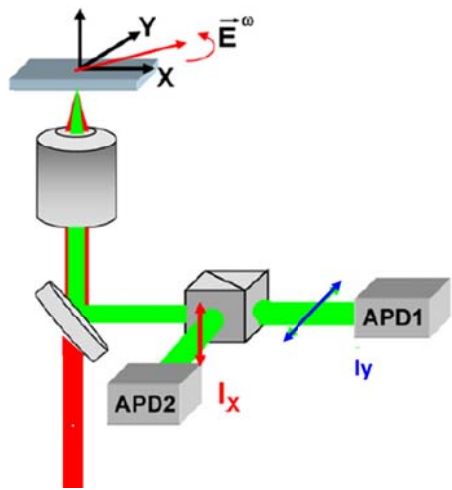

**Figure S1.** Experimental set- up. The sample is mounted on a piezo stage and scanned along both X and Y axes. The epi-reflected SH signal is collected by the same objective,

and its two perpendicular polarization components are detected by two calibrated avalanche photodiodes (APD).

**Symmetry consideration for the second order susceptibility**

Assuming a quasi-plane wave behavior at the focus of the confocal microscope with its beam-waist in the (X,Y) plane, one can define the fundamental electric field by its two in-plane components, $\vec{E} = (E_X, E_Y) = (E\cos\alpha, E\sin\alpha)$, where the field is taken as linearly polarized, with an angle defined with respect to the X axis and amplitude E. The sample is illuminated at the fundamental frequency, and the Second Harmonic Generated signal is collected through the same optical path, and its dependence on input field polarization is measured. Let us examine the polarization plots for two extreme cases: an object with threefold symmetry and an object with twofold symmetry.

    i)    A planar object with a threefold rotational symmetry

As discussed in the paper, for such an object the $\chi^{(2)}$ tensor is composed of two coefficients, $\chi^{(2)}_{YYY}$ and $\chi^{(2)}_{YXX}$ which are linked by relation (1), namely corresponding to $\rho=-1$ where $\rho$ is the nonlinear anisotropy parameter as defined in the paper. Under these conditions, define $\chi^{(2)}_{//} \equiv \chi^{(2)}_{YYY} = -\chi^{(2)}_{YXX}$ the single remaining independent coefficient of the $\chi^{(2)}$ tensor. The X and Y components of the induced SHG polarization vector can be expressed as a function of the incoming fundamental field by:

$$P_X^{2\omega} = -\chi^{(2)}_{//} \sin(2\alpha) E^2 \quad \text{and} \quad P_Y^{2\omega} = -\chi^{(2)}_{//} \cos(2\alpha) E^2$$

Approximating the far field intensities along X and Y as proportional to the squared components of the induced polarization along the corresponding axis leads to the following expressions:

$$I_X^{2\omega} \propto \left[\chi^{(2)}_{//}\right]^2 \sin^2(2\alpha)(I^\omega)^2 \quad \text{and} \quad I_Y^{2\omega} \propto \left[\chi^{(2)}_{//}\right]^2 \cos^2(2\alpha)(I^\omega)^2$$

These expressions account for the four-lobed cross-shaped pattern of the SHG polarization plot for a planar object abiding to three-fold symmetry. This possibly surprising four-lobed pattern emanating from a threefold symmetric object is the result of the symmetry breaking projection of the induced polarizations, and hence of the generated harmonic fields, along the X and Y polarization analysis direction. The more intuitive three-fold symmetry would appear if the object of interest were rotated while the polarizations of the input fundamental and generated harmonic would remain fixed and set parallel to each other. object

   ii)    A planar object with a twofold symmetry along its Y axis

Assuming the validity of Kleinman permutation symmetry The $\chi^{(2)}$ tensor under this symmetry is then composed of two independent coefficients $\chi^{(2)}_{YYY}$ and $\chi^{(2)}_{YXX} (= \chi^{(2)}_{XYX} = \chi^{(2)}_{XXY})$. The two independent coefficients are no more linked by relation (1), leading us to introduce the nonlinear anisotropy parameter $\rho \equiv \chi^{(2)}_{YXX} / \chi^{(2)}_{YYY}$. This anisotropy parameter reflects the deviation from three-fold symmetry, as would be the case if the individual equilateral triangles depicts in figure 2 in the paper were to couple to each other. For non-coupled entities, the threefold symmetry is maintained and ρ=-1, while for a fully coupled system, the $\chi^{(2)}_{YYY}$ completely dominates, and ρ=0. Thus, the stronger the coupling the closer is ρ to zero.

In the following derivations, we make use of only two parameters, namely

$$\chi^{(2)}_{//} = \chi^{(2)}_{YYY} \quad \text{and} \quad \rho, \text{ with } \chi^{(2)}_{YXX} = \rho \chi^{(2)}_{//}$$

The expressions for the induced polarizations are then

$$P_X^{2\omega} = \rho \chi^{(2)}_{//} E^2 \sin(2\alpha) \quad \text{and} \quad P_Y^{2\omega} = -\chi^{(2)}_{//} E^2 \left[ -\cos(2\alpha) + (1+\rho)\cos^2 \alpha \right]$$

Following a similar line of derivation as for the previous case, the harmonic intensities detected along X and Y are then respectively given by

$$I_X^{2\omega} \propto \rho^2 \left|\chi_{//}^{(2)}\right|^2 (I^\omega)^2 \sin^2(2\alpha)$$

$$I_Y^{2\omega} \propto \left|\chi_{//}^{(2)}\right|^2 (I^\omega)^2 \left[\cos^2(2\alpha) + (1+\rho)\left[(\rho-3)\cos^2(\alpha) + 2\right]\cos^2(\alpha)\right]$$

For a three-fold symmetric object (ρ=-1) these general expressions reduce to those of the previous case. In the limit of very strong coupling with ρ=0 the expressions take the following form

$$I_X^{2\omega} = 0 \text{ and } I_Y^{2\omega} \propto \left|\chi_{//}^{(2)}\right|^2 (I^\omega)^2 \sin^4\alpha$$

It is interesting to note that the expression of $I_X^{2\omega}$ maintains the same shape as in the three fold case as derived above (subject to a $\rho^2$ scaling factor). In particular, whatever the value of ρ, in the X analysis direction, the four-lobed shape that is characteristic of octupolar systems is maintained, although the three-fold symmetry is lost. Its amplitude decreases with $\rho^2$ down to the strong coupling case of ρ=0 where the 'memory' of the three-fold symmetry vanishes. This is in agreement with our experimental observations as in Fig. 6 in the paper.

Figure S2 depicts different harmonic polarization plots which display a progressive alteration when the apparent SHG response changes from a perfect four lobed shape for both X and Y polarizations in the non-interacting case (ρ =-1) onto a more dipolar shape for ρ =-0.5.

As shown in Figure S3, In the strong coupling limit, the polar plot takes a characteristic $\sin^4\alpha$ "eight"-like shape along the Y interaction axis, consistently with our experimental results.

Thus, by fitting the experimental polar plot data to this model, and extracting the value for ρ that best describes the data, we can discuss the degree of coupling between individual objects that is responsible for lowering of symmetry as expressed by the departure of the ρ factor from -1.

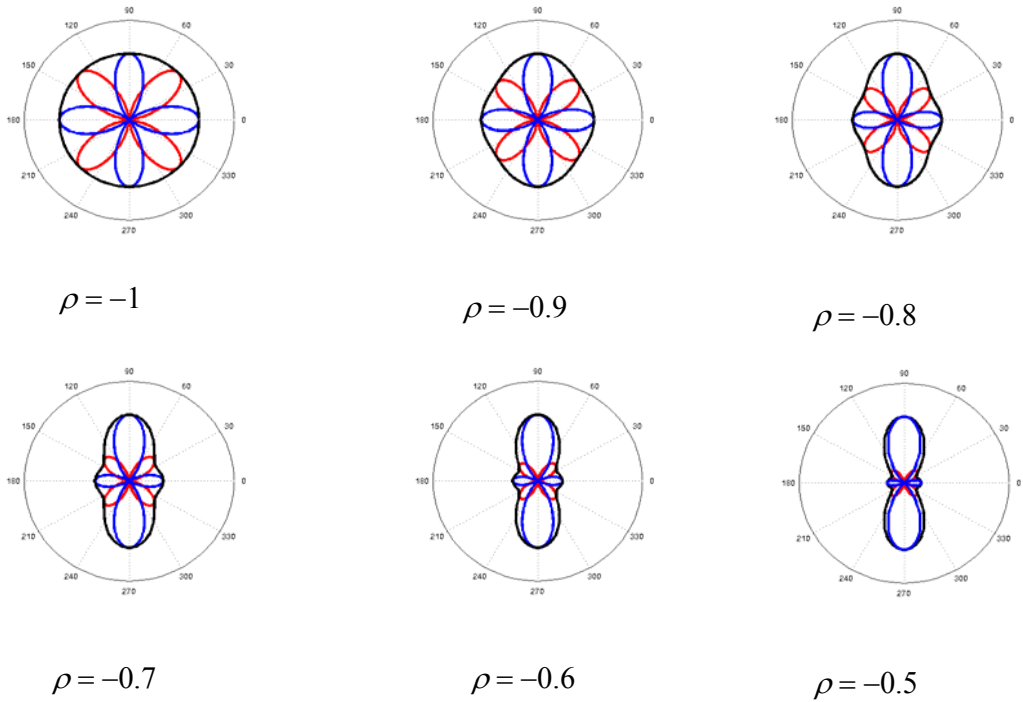

Figure S2: Calculated polar plots of the SHG emission pattern from a triplet of objects with three fold symmetry aligned along the Y axis as a function of the input beam polarization. The coupling between objects is expressed by the $\rho$ coupling parameter. When strong coupling occurs $\rho = 0$ and all the all dipoles are emitting in phase. When the objects do not interact, $\rho = -1$, and the emission is similar to that of an individual object. Blue curves: the Y-axis component of the generated SHG, red curves: the X-axis component and black curves is the total emission. Polarization plots are shown for different $\rho$ values ranging here from -1 to -0.5 (i.e.) from non interacting nano-objects with equilateral symmetry to a significant interaction level which lowers the three fold symmetry in the independent case. The simulations shown in this Figure do not include corrections for the not fully achromatic dichroic mirror and reproduce expressions in this part.

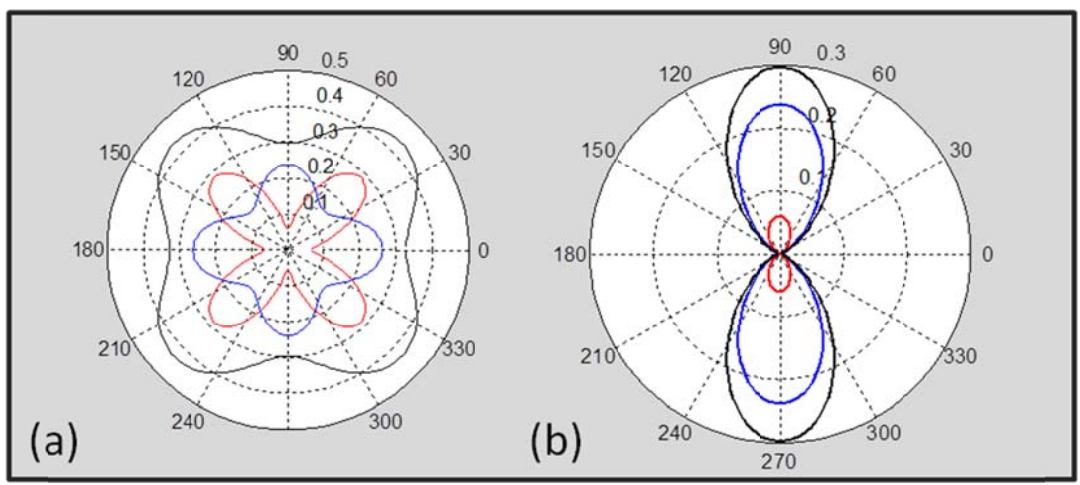

Figure S3: Same as above for two extreme cases, namely ρ =-1 (non interacting nano-objects) and ρ =0 (strongly interacting nan-objects), now including in both case corrections for the not fully achromatic dichroic mirror.